\documentclass[12pt]{article}
\usepackage{amssymb,graphicx}

\makeatletter
\renewcommand\section{\@startsection {section}{1}{\z@}%
                                 {-3.5ex \@plus -1ex \@minus -.2ex}%nn
                                   {2.3ex \@plus.2ex}%
                                   {\normalfont\large\bfseries}}
\renewcommand\subsection{\@startsection{subsection}{2}{\z@}%
                                   {-3.25ex\@plus -1ex \@minus -.2ex}%
                                     {1.5ex \@plus .2ex}%
                                     {\normalfont\bfseries}}
\renewcommand\subsubsection{\@startsection{subsubsection}{3}{\z@}%
                                   {-3.25ex\@plus -1ex \@minus -.2ex}%
                                     {1.5ex \@plus .2ex}%
                                     {\normalfont\itshape}}
\makeatother

\newcommand{\Letter}{
    \setlength{\textwidth}{16.5cm}
    \setlength{\textheight}{22.6cm}
    \hoffset=-0.6in
    \voffset=-2.1cm }

\Letter

\newcommand{\bra}[1]{\langle #1|}
\newcommand{\ket}[1]{|#1 \rangle}

\newcommand{\gsim}{ \lower .75ex \hbox{$\sim$} \llap{\raise .27ex \hbox{$>$}} }
\newcommand{\lsim}{ \lower .75ex \hbox{$\sim$} \llap{\raise .27ex \hbox{$<$}} }

\begin{document}
\thispagestyle{empty}
\begin{flushright}
\parbox[t]{1.5in}{
MAD-TH-05-5\\
SISSA-62/2005/EP \\
hep-th/0508229}

\end{flushright}

\vspace*{0.5in}

\begin{center}
{\large \bf Warped Reheating in Multi-Throat Brane Inflation}

\vspace*{0.5in} {Diego Chialva${}^{1}$, Gary Shiu${}^{2}$, Bret
Underwood${}^{2}$}
\\[.3in]
{\em
     ${}^{1}$ International School for Advanced Studies (SISSA)
     Via Beirut 2-4, I-34013 Trieste, Italy

     ${}^{2}$ Department of Physics,
     University of Wisconsin,
     Madison, WI 53706, USA
         }
\end{center}

\bigskip

\begin{center}
{\bf
Abstract}
\end{center}
\noindent  We investigate in some quantitative details the viability of reheating in multi-throat brane inflationary %%@
scenarios by estimating and comparing the 
time scales for the various processes involved.  
We also calculate within perturbative string theory
the decay rate of excited closed strings into KK modes and compare with that of their decay into gravitons; we find %%@
that in the inflationary throat the 
former is preferred.
We also find that 
 over a small but reasonable range of parameters of the background geometry,
these KK modes will preferably 
tunnel to another throat (possibly containing the Standard Model) instead of decaying to gravitons due largely
to their suppressed coupling to
the bulk gravitons.
Once tunneled, the same 
suppressed coupling to the gravitons again allows them to reheat the Standard Model efficiently.  
We also consider the effects of adding more throats to the system and find that for extra throats with small warping, %%@
reheating still seems viable.

\vfill

\hrulefill\hspace*{4in}

{\footnotesize
Email addresses: 
chialva@sissa.it, shiu@physics.wisc.edu, bjunderwood@wisc.edu.}

\newpage

\section{Introduction}

Inflationary models from string theory
have become more refined in the last few years \cite{DvaliTye,DD,StringInflation,DBI,KKLMMT,Cremades:2005ir}, 
due largely to the
observation that branes and fluxes can play an important role in early universe cosmology.
A key element in many of these explicit  constructions 
is the idea of brane inflation \cite{DvaliTye}, where the interaction 
between branes provides
a microscopic origin of the inflaton potential. To embed this idea in a full-fledged string 
theoretical model, however, there are
more hurdles to clear;
the most important of which is
the problem of moduli stabilization.
Moreover,
the mechanism which stabilizes moduli may add new constraints to this scenario and could a priori 
ruin the successes of brane inflation.
This and related issues were addressed 
in a concrete Type IIB string theory setting in \cite{KKLMMT}, 
utilizing the fact that
all geometric moduli can be stabilized by background fluxes %%@
\cite{Gukov:1999ya,Dasgupta:1999ss,Taylor:1999ii,Greene:2000gh,Curio:2000sc,GKP,Becker:2001pm,Kachru:2002he}
and non-perturbative effects \cite{KKLT}. 
Thus, a rather rich framework for 
inflation has emerged from our current, albeit still limited, understanding of flux compactification
in string theory.

A by-product of stabilizing moduli by fluxes is that the background geometry comes
naturally equipped with a warp factor. 
In particular, strongly warped regions, or ``warped throats",  with exponential warp factors
can arise when there are fluxes 
supported 
on cycles 
localized in small regions of 
the compactified space.
A prototypical  example of such a strongly warped throat
is the warped deformed conifold solution of Klebanov and Strassler \cite{KS,Klebanov:2000nc}.
Indeed, warping is ubiquitous in many string inflationary models, e.g.,  $D\overline{D}$ inflation \cite{KKLMMT}, 
tachyonic inflation \cite{Cremades:2005ir}, and DBI inflation \cite{DBI}. The reason is that
in addition to generating the hierarchy between the weak scale and the Planck scale  
as in the Randall-Sundrum scenario \cite{RS}, 
warping
can also help in flattening the inflaton potential and/or slowing down the D-branes
in order to obtain a sufficient number of e-foldings of inflation.  Even with warping, however, it appears that 
general contributions from stabilizing the K\"ahler moduli generate a large mass for the inflaton, the $\eta$-problem, %%@
although some investigations of this issue have already begun \cite{Tye, Sarah, Dynamical}.

Given that we now have a wide variety of 
string inflationary models with warped throats, a natural next step 
in this program is
to investigate the viability of reheating in these models (see \cite{Preheat} for a review of reheating). However, to %%@
examine the reheating 
problem in a concrete setting, we
need to understand better how the Standard Model can arise in flux compactification.
Fortunately, much progress has been made in the past year in constructing Standard Model-like
flux vacua \cite{Cascales:2003wn, Marchesano:2004yq,Marchesano:2004xz,Blumenhagen:2005mu}.
Thus, the time is ripe to visit, in more
quantitative terms,  the 
issue of warped reheating in the brane inflationary scenario \cite{Reheating,Kofman,Frey:2005jk}.

A detailed study of 
reheating in single throat brane inflation has recently been undertaken in \cite{Kofman}. 
While reheating seems viable in models with a single throat, there are several challenges that 
one needs to overcome in building fully realistic models:
(i) First of all, 
there is not much room in generating more than one hierarchy 
with a single throat\footnote{An exception is when 
there are
throats 
within throats so that the warp factor is discontinuous in the radial direction, see, e.g., \cite{Cascales:2005rj}.}.
Unless additional mechanisms of generating hierarchies (e.g., dynamical effects such as supersymmetry
breaking) are invoked,
it seems difficult to generate simultaneously
the weak scale hierarchy and a correct level of density perturbation
$\frac{\delta \rho}{\rho} \sim \frac{H}{M_{Pl}} \sim 10^{-5}$, at least within the context of
single field inflation.
(ii) In a generic KKLT-type model, at least its simplest version,
the scale of supersymmetry breaking is intimately tied to the inflationary scale \cite{Kallosh:2004yh}.
Although there are ways out of this conundrum \cite{Kallosh:2004yh}, such considerations impose rather strong %%@
constraints on model building.
(iii) On a technical side, since the Standard Model branes and the $D\overline{D}$ pair
which drives inflation reside in the same throat,
the energy resulting from the annihilation can partition among {\it both} the open and closed
string degrees of freedom. 
Unlike 
$D \overline{D}$
annihilation, the process of tachyon condensation in the presence of leftover D-branes,
though considered \cite{Leblond} is certainly less understood.

In view of these challenges, multi-throat brane inflation 
seems more appealing as
one can generate several
hierarchies of energy scales by the warp factors associated with different throats.
There is also no need for additional branes to reside in the throat where the
tachyon condenses. Hence the aforementioned problems with single throat inflation
can be solved in one stroke.
However, one should be cautious in applying this ``modular" approach
in model building, i.e., by introducing
separate throats to generate
widely different scales. As recently pointed out in \cite{Frey:2005jk}, if the warping differs 
significantly between throats, string and Kaluza-Klein (KK) effects can dramatically alter
the usual 4D effective field theory description of inflation and reheating. 
Nonetheless, irrespective of the warped scales they generate during inflation,
having more throats can certainly give us more leeway in building realistic models. 
Most importantly, since the cosmic strings formed at the end of brane inflation are spatially separated
from the Standard Model branes in multi-throat scenarios,
they are not susceptible to breakage \cite{Cosmic} and could survive cosmologically as a network
of cosmic F- and D-strings and give
rise to interesting signatures of string theory \cite{CosStrings,Cosmic,FandD,CosmicStringReviews}.
Thus, there are strong theoretical reasons to evaluate the viability of
this scenario.

At the face of it, multi-throat brane inflation appears to fail on the front of reheating because
there are no direct couplings between the inflationary and Standard Model branes.
The two throats communicate only weakly through gravity and given the energy barrier produced by
the warping of the bulk separating the throats, it seems to be a major challenge 
to successfully channel the energy from the $D\overline{D}$ annihillation into standard model degrees of freedom. 
Some preliminary studies of reheating in such multi-throat systems have been initiated
in \cite{Reheating, Kofman,Frey:2005jk}. 
In particular, it has been suggested in \cite{Reheating} that 
the KK modes of the graviton can efficiently reheat the Standard Model, since the wavefunctions of the KK modes are %%@
sharply peaked at 
the tip of the throats.  
However, reheating is a dynamical process with several time scales involved. It is not a priori obvious
if this interesting observation will be enough to guarantee efficient reheating when the full dynamics are taken into %%@
account. 
In particular, the issue of tunneling seems to be the most serious 
obstacle \cite{Reheating} for multi-throat reheating
%when the inflaton and the Standard Model are localized at different throats.
because the KK modes could instead decay to gravitons in the inflationary throat.
Therefore, whether the Standard Model is reheated in a
conventional way \cite{Reheating,Kofman} or via more exotic stringy phenomena \cite{Frey:2005jk},
we need to first establish that the post-inflation energy can be 
effectively channeled to the Standard Model
throat.

The purpose of this paper is to investigate
in some {\it quantitative details} the viability of reheating in multi-throat brane inflationary scenarios,
building on some observations made in Refs \cite{Reheating, Kofman,Frey:2005jk}.  In particular, we investigate the %%@
decay of the excited closed string 
end products of $D\overline{D}$ annihilation into KK modes and gravitons; we find that in flat space, the decay to %%@
gravitons is favored: the decaying closed strings prefer to change their oscillator level by a small amount, and since %%@
the KK modes have 
non-zero momentum in the radial direction of the throat, 
they have a suppression factor due to their reduced phase space.  
However, in a warped throat the couplings of KK modes to closed strings is enhanced by two powers of the warp factor, %%@
and even for small
warping this enhancement is enough that the KK modes are favored decay products.  
These KK modes can then either tunnel from one throat to the other or decay into gravitons; surprisingly, we find that %%@
over
a small but reasonable range of parameters of the background geometry the KK modes prefer to tunnel into another %%@
throat.  This is largely due to the suppressed coupling between the KK modes and the
graviton deep in the throat.  These modes, now localized in the Standard Model throat, will then
decay to Standard Model degrees of freedom, reheating the Universe, and we find that the reheating temperature can be %%@
in an acceptable range.  It was suggested in
\cite{Kofman} that adding additional throats may ruin reheating in this scenario, due mainly to the late-time decay of %%@
KK modes in the other throats; we find, 
however, that for mildly warped throats the decay from the KK modes in other throats reheats the Standard Model at an %%@
acceptable temperature.  We should emphasize
that while our analysis of warped reheating
has KKLMMT-like constructions in mind, the results are applicable to
various other warped inflationary models such as the scenarios of \cite{DBI}
and \cite{Cremades:2005ir}.

This paper is organized as follows. For completeness and to set up our notation,
we review in Section 2 the setup of 
throat geometry in Calabi-Yau compactifications.
In 
Section 3 we discuss the chain process in which the $D\overline{D}$ annihilate, decay into KK modes and gravitons %%@
within the throat, and tunnel out of
the inflationary throat.  In Section 4 we examine the decay of KK modes localized in the Standard Model throat into %%@
gravitons or Standard Model
degrees of freedom, and estimate the reheating temperature.
In Section 5 we discuss how additional throats will change our
analysis.  We conclude in Section 6. Some details are relegated to the appendix.

\begin{figure}[h]
\includegraphics[scale=.55]{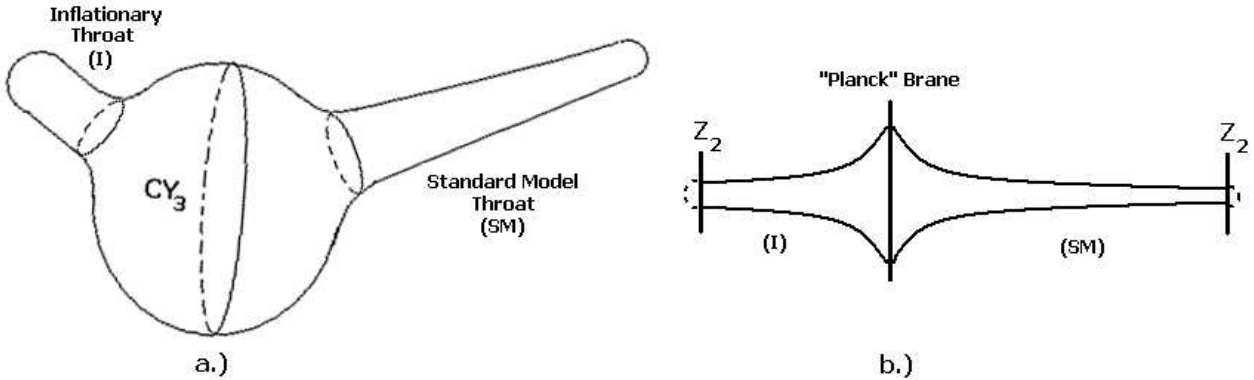}
\caption{{\small We will consider a KKLMMT setup with two throats, as in a.); as a simplification, we will consider 
the throats to be $AdS_{5}$, and glue the two throats together at a ``Planck" brane, as in b.).}}
\label{fig:2Throats}
\end{figure}

\section{Setup}
\setcounter{equation}{0}

The setup is a Type IIB compactification on a Calabi-Yau (CY) 3-fold with NS-NS and R-R fluxes turned on along the %%@
internal compact dimensions.
As in \cite{KS,Klebanov:2000nc},
by turning on fluxes on the cycles associated with a conifold one can generate a strongly warped ``throat'' which is %%@
glued to the bulk CY compact 
space.  The fluxes are quantized by:
\begin{eqnarray}
\frac{1}{2\pi\alpha'}\int_{A}F_{(3)} &=& 2\pi M \nonumber \\
\frac{1}{2\pi\alpha'}\int_{B} H_{(3)} &=& 2\pi K,
\label{eq:FluxQuant}
\end{eqnarray}
where $A$ and $B$ are the cycles on which the fluxes are supported.  The throat is a warped deformed conifold where %%@
the deformation replaces the conifold
singularity with an $S^{3}$ ``cap".  Far from the tip of the throat the geometry looks like an exact conifold with %%@
6-dimensional metric 
$ds^{2}_{6}= dr^{2}+r^{2}ds^{2}_{X_{5}}$.  The 5-dimensional space $X_{5}$ is some Einstein-Sasaki manifold (e.g., %%@
$T^{1,1}$)
whose details we will not be concerned with here.

Far from the deformation the throat can be described by the metric \cite{Kofman}:
\begin{equation}
ds^{2} = H^{-1/2}(r)g_{\mu\nu}dx^{\mu}dx^{\nu} + H^{1/2}(r)(dr^{2}+r^{2}ds^{2}_{X_{5}}),
\label{eq:ExactConifold}
\end{equation}
where the warping is,
\begin{equation}
H(r) = \frac{1}{r^{4}}(R_{+}^{4}+R_{-}^{4}\log(\frac{r}{R_{+}})^{4}).
\end{equation}
$R_{+}$ and $R_{-}$ are the radii of the $X_{5}$ at the top and bottom of the throat, respectively, and are given in %%@
terms of the flux quantizations
as:
\begin{eqnarray}
R_{+}^{4} &=& \frac{27\pi}{4}\ell_{s}^{4}g_{s}MK \nonumber \\
R_{-}^{4} &=& \frac{3}{8\pi}\frac{27\pi}{4}\ell_{s}^{4} g_{s}^{2} M^{2}.
\end{eqnarray}
The throat geometry can be further simplified as $AdS_{5} \times X^{5}$ where $R_{+}\approx R$ is the $AdS$ curvature %%@
scale and the hierarchy between
$R_{+}$ and $R_{-}$ sets the amount of warping in the throat $\frac{y_{0}}{R}\sim %%@
\frac{1}{4}\left(\frac{R_{+}}{R_{-}}\right)^{4}$, where the warp factor is approximately
\begin{equation}
h\equiv H^{-1/4}(r_{0})\approx \frac{r_{0}}{R_{+}} = e^{-y_{0}/R} = e^{-2\pi K/3Mg_{s}}.
\end{equation}

Using the coordinate transformation $r/R = e^{-ky}$ ($k\equiv R^{-1}$) one can put the equation of motion for the %%@
metric flucuation 
$g_{\mu\nu} = \eta_{\mu\nu}+h^{(n)}_{\mu\nu}(x,y,\rho_{i})$ (where $\rho_{i}$ are the coordinates associated with the %%@
$X_{5}$) in the form \cite{Kofman}
($h^{(n)}_{\mu\nu}(x,y,\rho_{i}) = e^{ip\cdot x}\chi^{(n)}(y)\Omega_{L}(\rho_{i})\epsilon_{\mu\nu}$, where %%@
$\epsilon_{\mu\nu}$ is the polarization):
\begin{equation}
(e^{4ky}\partial_{y}e^{-4ky}\partial_{y}+m^{2}e^{2ky}-k^{2}L^{2})\chi^{(n)}(y) = 0,
\end{equation}
where $m^{2}=p^{2}$ is the 4-D mass of the KK mode and $L^{2}$ is the quantized angular momentum on the $X_{5}$ whose %%@
structure we will leave
unspecified.  The solution of this equation is given by combinations of Bessel functions times an exponential:
\begin{equation}
\chi^{(n)}(y) = N e^{2ky}[J_{\nu}(m_{n}Re^{ky})+B Y_{\nu}(m_{n}Re^{ky})],
\label{eq:BesselSoln}
\end{equation}
where $\nu^{2}=4+L^{2}$.
(The bulk graviton zero mode is given as the limit as $m_{n}\rightarrow 0$).  The quantization of the KK masses can be %%@
seen by imposing orbifold boundary
conditions at the tip of the throat: $\partial_{y}\chi^{(n)}|_{y=y_{0}} = 0$ implies that $J_{\nu-1}(m_{n}Re^{ky_{0}}) %%@
=0$, so we find that the masses
are quantized in units of the zeros of the Bessel function, $m_{n}\sim \frac{n e^{-ky_{0}}}{R}$ for mode numbers $n %%@
\gg \nu-1$ (more precisely, \cite{Kofman}
find that KK modes localized near the bottom of the throat are quantized in units of the radius of the $S^{3}$ cap %%@
$m_{n}\sim \frac{n e^{-ky_{0}}}{R_{-}}$).  Imposing continuity and jump boundary conditions across the Planck brane as 
in \cite{Small}, we find $B\approx \frac{\pi}{8}\left(\frac{m_{n}}{k}\right)^{2}$\footnote{Note that 
the numerical coefficient is slightly
different from that used in \cite{SMCoupling3,RS} where the boundary
condition at the Planck brane was that the derivative of the wavefunction vanish, while in our case we merely require %%@
that the
derivative on either side of the Planck brane be equal and opposite (see Appendix).  Imposing different boundary %%@
conditions
at the Planck brane can possibly give different behaviors of the wavefunction at the tip of the throat, but it is %%@
unclear whether
these boundary conditions will lead to tunneling as in \cite{Small}.  We would like to thank Hooman Davoudiasl for %%@
clairifying this
point.}.

The string scale is related to the 4-D Planck mass by $M_{Pl}^{2} = M_{s}^{8}V_{6} g_{s}^{-2}\epsilon^{2}$, where %%@
$V_{6}$ is the volume of the Calabi-Yau 
compactification, which we take to be slightly larger than $R_{+}^{6}$, and $\epsilon = %%@
\left(\frac{2}{(2\pi)^{7}}\right)^{1/2}$. Because $V_{6}\approx\beta R_{+}^{6}$
where $\beta$ is various factors of $2\pi$ from integrating out the extra dimensions in the conifold, we will take %%@
$V_{6}\epsilon^{2}\approx R_{+}^{6}$.
In the paper we will be interested in staying within the SUGRA description
where we can ignore trans-stringy excitations, thus we wish to have $R_{-}^{-1} < M_{s}$.  We will parameterize the %%@
excess as 
$R_{-} M_{s}\approx R_{+}M_{s}/(\ln h_{i}^{-1})^{1/4} \approx \lambda (g_{s})^{1/4}/(\ln h_{i}^{-1})^{1/4}$,
where $\lambda = (\frac{27\pi}{4}MK)^{1/4}$ and we wish to keep track of the explicit dependence on $g_{s}$.  For deep %%@
enough throats almost all of
the flux in the quantization conditions in Eq.(\ref{eq:FluxQuant}) lies in the throats so one can have independent %%@
$AdS$ curvature scales for different
throats, but for simplicity we will consider both throats to have the same $AdS$ scale, so the product $MK$ must be %%@
the same in both throats.  (One can
see that allowing the $AdS$ scales to differ gives us extra parameters to vary).  We then have three 
parameters: $\lambda$ explained above, $h_{i}$ the warp factor of the I throat, and $h_{sm}$ the warp factor of the SM %%@
throat (we will take $g_{s}$ as 
fixed by requiring 
that $g_{s}\sim g_{YM}^{2} \sim \alpha$). These parameters can be further related by noting that the warp factor %%@
$h\equiv \frac{M}{M_{s}}$ gives the local string
scale $M$ in the throat; because $M_{s}$ is related to $\lambda$ through the definintion of the Planck scale, $h_{i}$ %%@
and $h_{sm}$ have implicit $\lambda$ dependence as well.  
Making this dependence explicit, 
\begin{eqnarray}
h_{i} &=& \frac{M_{i}}{M_{Pl}}\frac{\lambda^{3}}{g_{s}^{1/4}} \nonumber \\
h_{sm} &=& \frac{M_{sm}}{M_{Pl}}\frac{\lambda^{3}}{g_{s}^{1/4}}.
\label{eq:WarpLambda}
\end{eqnarray}
We will fix $M_{i}$ by requiring that the correct level of density perturbations $\frac{\delta \rho}{\rho}\sim %%@
\frac{H}{\epsilon M_{Pl}}$ is generated by the
brane-antibrane potential (which is set by the local string scale in the inflationary throat).  However, we will not %%@
commit ourselves to fixing the local string scale $M_{sm}$ at the tip of the Standard Model throat
to be TeV scale during inflation.
This will gives us more flexibility in satisfying the phenomenology of reheating and 
presumably 
other mechanisms can be used in conjunction with warping to generate the required hierarchy.
Another possibility is that the Standard Model throat relaxes to its ground state after inflation \cite{Frey:2005jk} %%@
because generic arguments suggest that
stringy corrections will tend to set $M_{sm}\sim H$ during the inflationary era,
so a throat with a local string scale less than $H$ will generically become shorter during inflation after
these corrections are taken into account.

\section{Chain Processes}
\setcounter{equation}{0}

The reheating of the Standard Model is the last step of a sequence of processes that starts in the inflationary throat 
through $D\overline{D}$ annihilation and, through a chain of decays and tunneling, ends when the inflaton energy
is transferred to the Standard Model degrees of freedom.  The steps in the chain decay are:

 \begin{itemize}
 \item the transfer of energy from the tachyon to closed strings at the end of inflation (Section 3.1),
 \item the subsequent decay of closed strings into massless string states (Kaluza-Klein modes and gravitons) (Section %%@
3.2),
 \item the interactions among these lower energy degrees of freedom (Section 3.3),
 \item the transfer of energy into the other throat(s) (via tunneling) (Section 3.4),
 \item the decay of these modes into Standard Model degrees of freedom and gravitons (Section 4).
 \end{itemize}

Throughout these chain processes, one must consider various cosmological constraints in order to determine
if a viable model of reheating can be constructed in a self-consistent way. 
For example, an overproduction of gravitons  
or the presence of relics could interfere with nucleosynthesis or
overclose the universe and must be avoided.
In the following we will analyze these issues systematically.

\subsection{$D\overline{D}$ annihilation} \label{DDannihilation}

The multi-throat scenario allows us to have a simple $D\overline{D}$ inflationary system in one throat and the %%@
Standard Model in a 
separate throat, sidestepping the problems of single throat inflation discussed in the introduction. Thus, the process %%@
of reheating is then more clearcut than a $DD\overline{D}$ system (single-throat model). The end of inflation
is triggered by the condensation of the open string tachyon
which develop when the distance between the brane-antibrane pair is sub-stringy,
leading to what is known as the tachyon matter \cite{Sen}. The latter can be thought of as 
a coarse-grained description of
the actual physical state, a distribution of closed strings \cite{Tachyons,Sen,Liu,Shelton}. 
By contrast, in the $DD\overline{D}$ case, the 
tachyon can couple directly to the open strings on the remaining (Standard Model) D-branes
instead of the usual process of decaying into closed strings. 
The reheating process is then be more complicated and less understood (see, however, \cite{Kofman} for comments about %%@
the suppression of the coupling to the 
open string sector).

The properties of the annihilation process into closed strings (the spectrum, the energy density and its distribution) 
can presumably be captured by the  
decay of unstable D-branes: the tachyon matter is an (asymptotically) pressureless gas with energy momentum components %%@
only in the directions orthogonal 
to the D-branes \cite{Tachyons,Sen,Liu,Shelton}.

The total average number density and the total average energy density emitted were found in \cite{Liu}:
 \begin{equation}
 \frac{\overline{N}}{V}=\sum_{s} \frac{1}{2 E_{s}}|A_{s}|^2~, \qquad  \frac{\overline{E}}{V}=\sum_s %%@
\frac{1}{2}|A_{s}|^2,
\label{eq:ClosedDensity}
 \end{equation}
where $V$ is the volume of the branes, $A_{s}$ is the amplitude for producing a state $s$, $E_{s}=\sqrt{k^2+N}$ is its %%@
energy and $N$ is the oscillator level (note that we are taking $\alpha'=4$).

For large $N$, $|A_{s}|^2\sim e^{-2\pi E_{s}}$ and the probability for every single state to be produced is
 \begin{equation}
 p_{s}=\frac{1}{2 E_{s}}|A_{s}|^2\sim \frac{1}{\sqrt{k^2+N}}e^{-2\pi \sqrt{k^2+N}}.
 \end{equation}
Notice that it depends only on the level number $N$ and not on the partition of $N$ among the various oscillators %%@
forming the state, so every state at a given 
level $N$ is equally produced. Moreover, for fixed $N$ the probability is peaked on the states that have %%@
$\frac{k^2}{N}\sim 0$, i.e. the closed 
strings produced will predominately have no Kaluza-Klein modes or winding modes in the Calabi-Yau space.

Furthermore, for the energy density,
 \begin{equation}
 \frac{\overline{E}}{V}=\sum_s \frac{1}{2}|A_{s}|^2= \sum_N \int d^{d-3}k\ D(N)e^{-2\pi \sqrt{k^2+N}}\sim \sum_N %%@
\epsilon,
 \end{equation}
where $D(N)$ is the degeneracy of states at level $N$ and $\epsilon$ is independent of $N$, we see that the set of %%@
states at every level number $N$ 
receives the same amount of energy density. In other words, 
energy is equipartitioned among the oscillator levels labeled by $N$.
If the maximum energy available for a single state is $\sim \frac{1}{g_s}$, 
then the maximum allowed level number is 
$N\sim \frac{1}{g_{s}^{2}}\sim 100$ if $g_{s}\sim .1$.

\subsection{Decay of Closed Strings into KK modes}
The next step in the energy conversion process is the decay of the closed strings into lighter degrees of freedom, %%@
namely the massless graviton and 
KK modes (see \cite{Kofman} for additional discussion). 
The Kaluza-Klein modes are localized deep inside the throat due to the warp factor, whereas the graviton has a %%@
non-zero wavefunction all 
over the Calabi-Yau. After proper normalization one finds that the KK mode wavefunctions are peaked deep in the throat %%@
by a 
factor of $h_{i}^{-1}$ relative to the graviton, so decay to the former is enhanced by a factor $h_{i}^{-2}$.  %%@
However, decay
to KK modes will be suppressed by phase space since the KK mass can be comparable to the local string scale, so it is %%@
not 
immediately clear whether KK modes or gravitons are preferred decay products.

We can estimate the decay rates into graviton and into Kaluza-Klein modes by an explicit 
(flat space) worldsheet computation. For a flat space computation to be valid 
the geometry must vary sufficiently slowly with respect to the string scale, i.e. for an AdS throat the AdS
length must be less than the string length.  However, since we require $R_{-}>\ell_{s}$ in order to use a SUGRA %%@
approximation,
$R_{+}>R_{-}>\ell_{s}$, where $R_{+}$ is the AdS length, this condition is satisfied. 
The geometry induces a potential that localizes the string,
and so the string effectively lives inside a box \cite{FandD}.  It must be stressed that since our calculation is for %%@
flat
space it will not capture all of the features of decay in curved space; however, as an estimate
we can 
compute the decay rates in flat space and convolve the results with the wavefunction square
of the corresponding decay products, i.e., graviton and KK modes respectively, in order to take into account the %%@
enhancement
of the KK mode wavefunction in the AdS space. 

Let us compute the average decay rate for closed strings fixing: (i) the mass of the initial state ($M^{2}=N$), (ii) %%@
one of the final states (the graviton or the Kaluza-Klein mode),
and (iii) the mass of the other final states ($M'^{2}=N'+p^2+k^2$).
We do not fix, however, the specific initial state (over which we average) or final state, other than specifiying its %%@
mass level.

 The decay is given by
 \begin{equation}
 \Gamma =g_{s}^{2}\int dP \sigma_{R}\ \sigma_{L} \nonumber
 \end{equation}
 where $d$ is the number of spatial dimensions, $\sigma_{R}$ is defined as
 \begin{equation}
 \sigma_{R}= \frac{1}{\mathcal{N}_{N_R}}
 \sum_{\Phi_{N}} \sum_{\Phi_{N'}}\langle \Phi_{N}| V_{R}(k,1)^{\dagger} |\Phi_{N'}
 \rangle \langle\Phi_{N'}|V_{R}(k,1) | \Phi_{N} \rangle \nonumber
,
 \end{equation}
the vertex operator we need is $V=V_RV_L$, with
 \begin{equation}
 V_{R}(p,z)=c(z)e^{ip\cdot X_{R}(z)}(i\xi\cdot\partial
 X(z)+\frac{\alpha'}{2}p\cdot\psi\xi\cdot\psi), \nonumber
 \end{equation}
$\int dP$ is the integral over the phase-space 
and $\mathcal{N}_{N_R}$ is the number of states at level $N=N_R=N_L$.

In particular, as we saw in Section (\ref{DDannihilation}), the closed strings produced when the $D\overline{D}$ %%@
system decays have preferably momentum, 
Kaluza-Klein and winding charges equal to zero. The result of the computation (see, e.g., \cite{CIR}) %%DC
can be written in terms of $N$ and $N_{0} = N-N' = 2 E \sqrt{N}$:
\begin{equation}
\cases{
\Gamma_{g} \sim g_{s}^{2}\pi \Omega_{3} \left(\frac{N_{0}}{2\sqrt{N}}\right)^{3} 
{ \frac{e^{-2a  (\sqrt{N} - \sqrt{N-N_0})}}{\left(1-e^{- \frac{a N_{0}}{2\sqrt{N-N_{0}}}}\right)^{2}}} & for graviton %%@
\cr
\Gamma_{KK} \sim g_{s}^{2} \pi\Omega_{3} \left(\frac{N_{0}}{2\sqrt{N}}\right)^{2} 
 \frac{e^{-2a  (\sqrt{N} - \sqrt{N-N_0})}}{\left(1-e^{- \frac{a N_{0}}{2\sqrt{N-N_{0}}}}\right)^{2}} %%@
\sum_{n=1}^{n_{max}} 
\sqrt{\frac{N_{0}^{2}}{4N}-k_{KK}^{(n)}} & for KK modes,\cr}
\label{eq:DecayRates}
\end{equation}
where $a=2 \sqrt{2} \pi$, $\Omega_{3}$ is the angular integral in three dimensions, and $k_{KK}^{(n)}$
is the Kaluza-Klein 
momentum of the $n$-th KK mode.  We have implicitly assumed that the KK mode has momentum only 
in the radial direction of the throat, although it is straightforward to 
include momenta in other directions of the Calabi-Yau as well.  Note that in order to get these results we have %%@
approximated the 
distribution of states $D(N)$ by its limit for large $N$, but actually this approximation
is already very good for moderate values of $N$.  

\begin{figure}[t]
\begin{center}
\includegraphics[scale=.3]{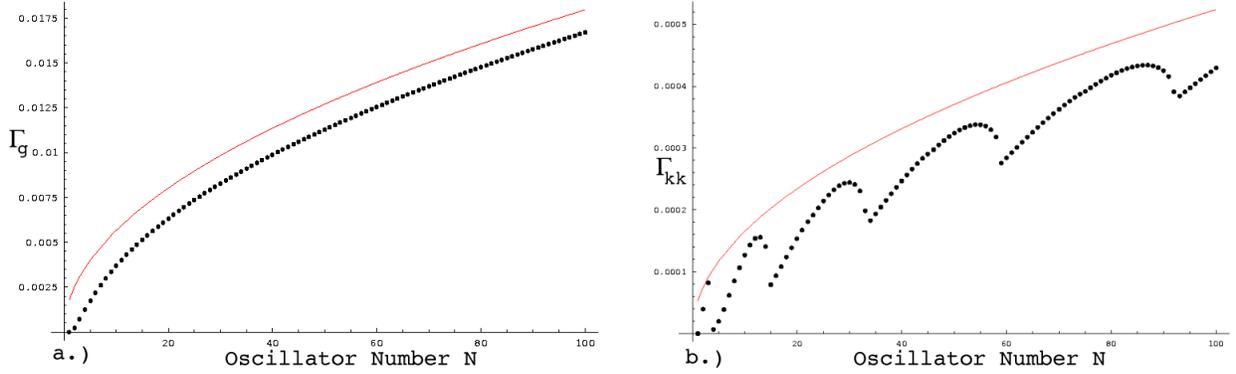}
\caption{\small The numerical value of the decay rate (in units of the local string scale) of a closed string into an %%@
a.) graviton and b.) KK mode, 
for all allowed processes, 
is plotted for finite oscillator level $N$ (black dotted line), which is relevant for the decay of the closed string %%@
end products from $D\overline{D}$ annihilation, 
and in the limit $N\rightarrow \infty$ (red solid line), which corresponds to the field theory limit of \cite{CIR}, %%@
where $\Gamma \propto \sqrt{N}$ (note that we have not included any warp
factor enhancement yet).  The jumps in b.) are from threshold effects for the production of KK modes; see
the discussion below Eq.(\ref{eq:GraviDecayLimit}).}
\label{fig:SumIntegral}
\end{center}
\end{figure}

To obtain the total decay rate into gravitons (or similarly KK modes) for any given initial state $N$, we sum over all %%@
the
allowed processes labeled by $N_0$ where $N_0=1$ to $N$.
Unlike \cite{CIR}, we have not taken $N \rightarrow \infty$ so the total
decay rate into gravitons $\Gamma_{g}$ is valid for an initial state of finite mass.
Recall that in the $N \rightarrow \infty$ limit, 
we can turn the sum over $N_{0}$ into an integral 
through the Jacobian $dN_{0} = 2 \sqrt{N}d\omega$, where $\omega =  N_{0}/(2 \sqrt{N})$,
\begin{equation}
\Gamma_{g, tot}\sim g_{s}^{2}\pi \Omega_{3} 2 \sqrt{N} \int_{0}^{\infty} d\omega\ \omega^{3} 
\frac{e^{-2 a \omega}}{\left(1-e^{- a \omega} \right)^{2}} = c_{g} \sqrt{N},
\label{eq:GraviDecayLimit}
\end{equation}
where $c_{g}\approx 0.0018$.  This corresponds to the field theory limit since as $N \rightarrow \infty$,
the mass of the initial state $M \rightarrow \infty$ or equivalently $\alpha' \rightarrow 0$.
As can be seen from Figure \ref{fig:SumIntegral},
taking finite $N$ gives
a smaller value for $\Gamma_{g,total}$ than would be expected from field theory,
however the parametric dependence of $\Gamma_{g}(N)\propto \sqrt{N}$ as discussed in \cite{CIR}
is unchanged. The difference between the two curves in Figure \ref{fig:SumIntegral}
is due to the stringy corrections to the decay rate, namely that decays must occur in discrete steps of energy.
One can obtain similar looking expressions for the KK modes, where once again the parametric dependence on $N$ is
$\Gamma_{KK}(N)\approx c_{KK} \sqrt{N}$, where $c_{KK}\approx 0.0002$ without the warp factor.  The finite $N$ %%@
solution shows some qualitatively different 
features than the limiting case:
the jumps are produced
when, for fixed $N_{0}$, $N$ becomes large enough that a KK mode cannot be produced, i.e. $n_{max}<1$.  In  
particular, the jump at $N=15$ is for $N_{0}=2$ (where $\lambda\sim 5$ and $g_{s}\sim .1$, which
as we will see below are reasonable values): above $N=15$ one cannot produce a KK mode with $N_{0}=2$, so we 
drop below threshold.  There are multiple jumps because we are summing over all $N_{0}$, so we pick up 
multiple threshold effects.  Notice how this is different from the $N\rightarrow \infty$
limit, where we only obtain the enveloping behavior.

The sum over the mode number $n$ of the KK modes of the square root in Eq.(\ref{eq:DecayRates}) is the phase space %%@
available to 
the decay when the KK mode has a non-zero 4D mass, and the sum is
 over all kinematically allowed KK modes.  Notice that for small $N_{0}$ the energy released by the decay is also %%@
small so it is difficult to produce KK modes, and 
the phase space factor reduces the rate of decay into KK modes.  For large $N_{0}$, i.e. $N_{0}\sim N$, more KK modes %%@
can be produced, 
however the decay rate then depends exponentially on $\sqrt{N}$ and is
highly suppressed.  
The phase space factor and the
exponential suppression\footnote{This exponential suppression is due to the sum over the oscillators part of the %%@
string states. 
Since it depends only on the local properties of the compactification, it should be present both in flat and curved %%@
spaces.} of decay with large $N_0$ compete to maximize
the decay rate into KK modes for fixed $N$ for $N_{0,max}^{KK}\sim a_{KK} \sqrt{N}$, where one can numerically
determine $a_{KK}\sim .65$,
so we see that only low lying KK modes ($n_{max}\sim \frac{.65\lambda g_{s}^{1/4}}{(\ln h_{i}^{-1})}$) 
are produced\footnote{Note that this also means that angular KK states, which are quantized at a higher level, are not %%@
produced in copious amounts so the problem discussed in \cite{Kofman} may not be present.}.  Similarly, one can %%@
determine
that gravitons are produced most often for $N_{0,max}^{g}\sim a_{g}\sqrt{N}$, $a_{g}\approx \frac{3}{2a}$.

\begin{figure}[t]
\begin{center}
\includegraphics[scale=.5]{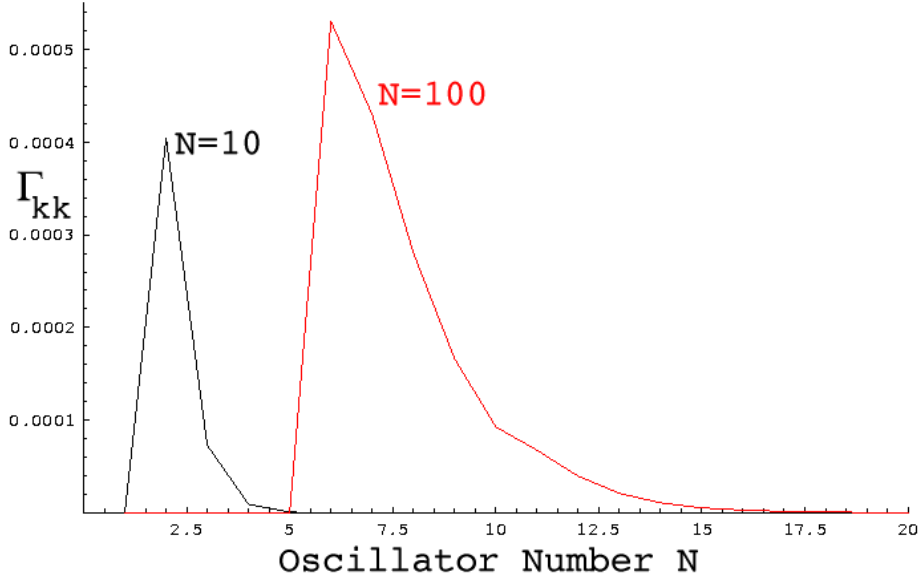}
\caption{\small The numerical value of the decay rate of a $N=100$ (red) and $N=10$ (black) closed string into a KK %%@
mode for 
different decay processes $N_{0}=N-N'$.  Notice that the decay rate
is peaked at small $N_{0}$ and falls off quickly, but increasing $N$ increases the $N_{0}$ at which the rate is %%@
maximum.  
The large initial jump when $N_{0}\approx N_{0,max}$ is due to the decay process crossing the threshold for production
of a KK mode.  One can show that $N_{0,max}\sim .65 \sqrt{N}$.}
\label{fig:KKDecaySteps}
\end{center}
\end{figure}

Without any enhancement from the warp factor we find that $\Gamma_{g}(N)/\Gamma_{KK}(N)\sim c_{g}/c_{KK}\sim .1$ over %%@
$N=1$ to $100$, and
so we see that a warp factor enhancement may help in making KK modes a preferred decay product; however, the order of
magnitude of the KK decay rate is strongly dependent on the size of the $S^{3}$ cap $R_{-}$ because this determines %%@
how easily a KK mode can
be produced.  Changing the value by a small amount leads to large variations in the decay rate, e.g. changing $R_{-}$ %%@
by a factor of 5 can modify
the KK decay rate by $\sim 7$ orders of magnitude ($c_{KK}\rightarrow 10^{-11}$)!  

What we are really interested in, however, is the ratio of the total amount of energy deposited into KK modes versus %%@
gravitons.  This can
be estimated through the following argument.  The typical time for an initial state $N_{in}$ to decay into either a %%@
graviton or KK mode is:
\begin{equation}
\Delta t_{N_{in}} \sim \frac{\Delta N}{\sum_{N_{0}=1}^{N_{in}} \Gamma_{g}(N_{in},N_{0})N_{0} + %%@
\Gamma_{KK}(N_{in},N_{0}) N_{0}},
\label{eq:FixedNTime}
\end{equation}
where $\Delta N \sim a_p \sqrt{N}$ is the typical step size. We will take
$a_p \sim a_{KK}$ if the KK modes have a larger decay rate
and $a_p \sim a_g$ otherwise.
The total number of species $i=\{\mbox{graviton, KK}\}$ produced by a closed string $N_{in}$ is then,
\begin{equation}
{\cal N}_{N_{in}}^{i}=\sum_{N=1}^{N_{in}} \left(\sum_{N_{0}} \Gamma_{i}(N,N_{0})\right)\Delta t_{N}\sim %%@
\sum_{N=1}^{N_{in}} \frac{a_p c_{i}}{a_{g}c_{g}+a_{KK}c_{KK}},
\label{eq:NumbDensN}
\end{equation}
where we have approximated $\sum_{N_{0}}\Gamma_{i}(N,N_{0})\sim c_{i}\sqrt{N}$ and $N_{0,max}^{i}\sim a_{i}\sqrt{N}$.  %%@
The total number of species
produced by the closed strings is then the weighted sum over all oscillator levels:
\begin{equation}
{\cal N}_{tot}^{i} = \sum_{N_{in}=1}^{g_{s}^{-2}} \sum_{s} p_{s}(N_{in}) {\cal N}_{N_{in}}^{i} = %%@
\sum_{N_{in}=1}^{g_{s}^{-2}} \sum_{N=1}^{N_{in}} 
\frac{a_p c_{i}}{(a_{g} c_{g} + a_{KK} c_{KK})\sqrt{N}};
\end{equation}
similarly, the fraction of energy carried by species $i$ is
\begin{equation}
E_{i} \sim \sum_{N_{in}=1}^{g_{s}^{-2}} \sum_{s} p_{s}(N_{in}) {\cal N}_{N_{in}}^{i} \frac{N_{0,max}}{2\sqrt{N}} \sim %%@
\sum_{N_{in}=1}^{g_{s}^{-2}} \sum_{N=1}^{N_{in}}
\frac{a_p a_{i} c_{i}}{(a_{g} c_{g} + a_{KK} c_{KK})2 \sqrt{N}}.
\end{equation}
We see then that the fraction of energy carried by KK modes versus gravitons is 
$E_{KK}/E_{g}\sim a_{KK}c_{KK}/(a_{g}c_{g})\approx 0.2 h_{i}^{-2}$, so even for mildly warped
throats $h_{i}\sim 10^{-2}$ we see that the KK modes appear to be preferred decay products.

One can estimate the total time for the decay of the closed strings into KK modes and gravitons by considering the %%@
decay time for the most massive closed string.
In this case, the total time for decay is the sum of each of the steps,
\begin{equation}
\Delta t_{tot} \sim \sum_{N=1}^{g_{s}^{-2}} \Delta t_{N} = \sum_{N=1}^{g_{s}^{-2}} \frac{a_p}{(a_{g} c_{g}+a_{KK} %%@
c_{KK})\sqrt{N}}.
\end{equation}
For the case where gravitons dominate, $a_{g} c_{g}\gg a_{KK} c_{KK}$, we have $\Delta t_{tot}\sim %%@
\frac{1}{c_{g}}\sum_{N=1}^{g_{s}^{-2}} \frac{1}{\sqrt{N}}$ 
$\sim 5 \times 10^{3} \sqrt{\alpha_{i}'}$, while for the case where the KK modes are dominant decay products due to %%@
the warp factor enhancement, 
$\Delta t_{tot}\sim \frac{1}{c_{KK}} \sum_{N=1}^{g_{s}^{-2}} \frac{1}{\sqrt{N}} = 5 \times 10^4 \sqrt{\alpha_{i}'}$.

\subsection{KK mode interactions}

Because the KK modes carry conserved quantum numbers they cannot decay into lighter degrees of freedom unless they %%@
collide with another
KK mode with opposite quantum number in the throat.  The time scale of such collisions is found by estimating when the %%@
average number of collisions
(the number density of the particle times the column depth times the cross-section of interaction) is approximately %%@
one, i.e.
$\tau\sim (n\sigma v)^{-1}$, where $n$ is the particle number density,
$\sigma$ is the cross-section, and $v$ is the velocity.  The KK modes as {\it initial} decay products
of excited closed strings are relativistic. This can be seen as follows. 
The average energy per KK mode is approximated by the average energy per closed string
 \begin{equation} \label{averageKKenergy}
 \bar E_{KK}\sim M_{s}h_{i}=M_{i}
 \end{equation}
and their mass is
 \begin{equation}
 m_{KK}\sim \frac{h_i}{R_-}
 \end{equation}
therefore $\frac{\bar E_{KK}}{m_{KK}}\sim \frac{R_-}{\sqrt{\alpha'}}$
and so we can assume $v\sim 1$ (taking $v<1$ will only increase the annihilation time).  
Moreover, as we will see, the KK modes thermalize quickly so it is reasonable to take $v \sim 1$
in estimating the time scales for the processes involved in the inflationary throat.
The KK states can annihilate and pair produce gravitons or other (kinematically allowed) KK modes.
The cross-section for KK modes to annhilate into bulk gravitons is set by the Planck scale 
$\sigma\sim M_{Pl}^{-2}$.  We will approximate the number density after inflation by the total energy 
density after inflation divided by the average energy per KK mode; the energy density after inflation is,
\begin{equation}
\epsilon_{I}=2T_{3}h_{i}^{4} \sim \frac{(M_{s}h_{i})^{4}}{g_{s}}=\frac{M_{i}^{4}}{g_{s}},
\label{eq:InflationEnergyDensity}
\end{equation}
where $T_{3}$ is the tension of a $D3$.  The average energy per KK
mode is, as we said above,  (\ref{averageKKenergy}).  The number 
density after inflation then is given by $n_{I}\sim \frac{M_{i}^{3}}{g_{s}}$.  Plugging these values in we can obtain %%@
a rough estimate for the
timescale for KK modes to decay to gravitons:
\begin{equation}
\tau_{g_{\mu\nu}}\sim g_{s}\left(\frac{M_{Pl}}{M_{i}}\right)^{2}\frac{1}{M_{i}}.
\label{eq:KKGravitonI}
\end{equation}

As mentioned above, KK modes can interact among themselves and thermalize, coming to a thermal temperature $T_{KK}$ %%@
(see also \cite{Kofman}).  
The timescale for thermalization
is similar to the timescale for KK modes to interact and decay to gravitons, except that the cross-section is now set %%@
by the
local scale, $\sigma\sim g_{s}^{2} (M_{s}h_{i})^{-2}=g_{s}^{2} M_{i}^{-2}$.  We immediately see, then, that the %%@
timescale for thermalization of the KK modes is
much smaller than the timescale for KK modes to decay into gravitons:
\begin{equation}
\tau_{thermal}\sim \frac{1}{g_{s}M_{i}}.
\label{eq:KKThermalI}
\end{equation}

We can estimate the thermal temperature of the relativistic KK modes by setting the thermal energy density equal to %%@
the initial
energy density of the $D\overline{D}$ system:
\begin{equation}
\epsilon_{therm} = g_{KK}T^{4}_{KK} = \epsilon_{I}\sim \frac{(M_{i})^{4}}{g_{s}}.
\label{eq:ThermDensity}
\end{equation}
The number of KK degrees of freedom $g_{KK}$ is approximately equal to the number of relativistic KK modes, as we will %%@
argue below.  The temperature
of the KK modes, then, is:
\begin{equation}
T_{KK}\sim \frac{M_{i}}{(g_{KK}g_{s})^{1/4}}.
\label{eq:KKtemp}
\end{equation}
Since only KK states with $m_{n}<T_{KK}$ are relativistic, we see that only states with mode numbers $i\leq %%@
i_{rel}\sim  (\frac{\lambda}{(\ln h_{i}^{-1})^{1/4}})^{4/5}$
are relativistic; since the number density of non-relativistic modes has an $e^{-m/T}$ suppression, most KK modes will %%@
occupy relativistic degrees of
freedom, thus $g_{KK}\approx i_{rel}$. Note that the actual low-lying zeros of the Bessel functions of order 2 do not %%@
follow the simple 
relation $m_{n}\sim n\frac{h_{i}}{R}$, but
instead are at a higher mass level than this relation would suggest.  As we will see below we expect $\lambda$ to be %%@
small, thus $i_{rel}$ will also be small, so only
the very lowest lying KK states will be excited. Since angular states, as discussed in \cite{Kofman}, are zeros of %%@
Bessel functions of a larger order $\nu^{2}=4+L^{2}$ 
where $L$ is a conserved quantum number of the internal space, the lowest lying angular KK modes will be above the %%@
small thermal temperature and thus we do not expect that they
will be present in significant numbers after thermalization.

\subsection{KK tunneling}

\begin{figure}[h]
\begin{center}
\includegraphics[scale=3]{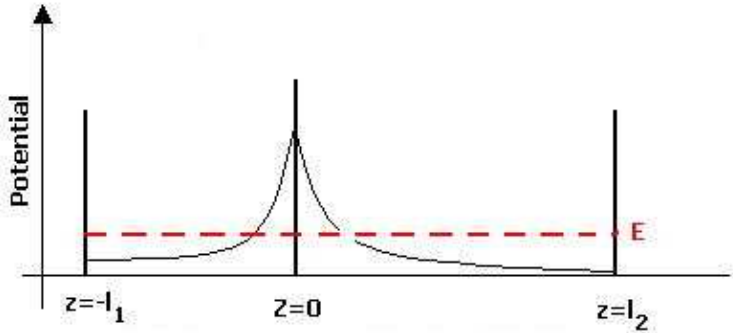}
\caption{{\small The warping creates a potential barrier for Kaluza-Klein states of the graviton in the Schr\"odinger %%@
coordinate system.  States initially localized in the inflationary (left) throat must tunnel through this barrier to %%@
communicate with the Standard Model (right) throat.  The Schr\"odinger energy of a state is $E=m^{2}$.}}
\label{fig:2ThroatsPotential}
\end{center}
\end{figure}

We would now like to estimate the time required for modes localized in the I throat to tunnel into the SM throat.  One %%@
can consider a toy model of the two throat
system by gluing two finite RS spaces together at the Planck brane as in \cite{Small,Reheating} (see Figure %%@
\ref{fig:2ThroatsPotential}).  
Under the coordinate transformation $z+R=R e^{ky}$, the metric can also be written in the form,
\begin{equation}
ds^{2}=\frac{R^{2}}{(|z|+R)^{2}}(\eta_{\mu\nu}dx^{\mu}dx^{\nu}+dz^{2}),
\label{eq:conformalMetric}
\end{equation}
where $R=R_{+}$ as before is the $AdS$ curvature scale.  We will place the IR brane for the inflationary throat at %%@
$z=-z_{i}\approx -R h_{i}^{-1}$ and the 
IR brane for the SM throat at $z=z_{sm}\approx R h_{sm}^{-1}$.  The utility of expressing the metric in these %%@
coordinates is that the equation of motion for the KK modes of the graviton,
\begin{equation}
h_{\mu\nu}(x,z)=\sqrt{\frac{R}{|z|+R}}e^{ip\cdot x}\psi_{\mu\nu}(z),
\label{eq:KKmode1}
\end{equation}
becomes a simple Schr\"odinger equation for $\psi$ (suppressing the indicies):
\begin{equation}
\partial_{z}^{2} \psi (z)+(m^{2}-\frac{15}{4(|z|+R)^{2}})\psi(z) = 0,
\label{eq:eqofmotion1}
\end{equation}
where the potential felt by the KK mode $V(z) = \frac{15}{4(|z|+R)^{2}}$ is proportional to the square of the warp %%@
factor (we have ignored the extra part of the potential
felt by the angular states).

The general solution to this equation of motion can be expressed in terms of Bessel functions,
\begin{equation}
\psi(z)=\sqrt{m(|z|+R)}[A J_{2}(m(|z|+R))+B Y_{2}(m(|z|+R))].
\label{eq:generalsoln1}
\end{equation}

We impose $\mathbb{Z}_{2}$ boundary conditions at the IR branes (which gives us our mass gap as discussed in Section %%@
2) and the appropriate Israel jump conditions at the Planck brane.
We would like to consider tunneling of an incoming state from the left hand throat into a state on the right hand %%@
throat for which \cite{Small} finds a tunneling probability for $m_{n}R\ll 1$ (certainly the case after thermalization %%@
for mode number $n\lesssim h_{i}^{-1}\sim 10^{4}$) of
\begin{equation}
P\sim (m_{n}R)^{4}.
\label{eq:TunnProb}
\end{equation}
In the calculation of \cite{Small} the reflection of the KK mode at the wall of the finite SM throat was not included, %%@
and one may worry that this reflection
may significantly alter the tunneling results.  However, one can see that the $\mathbb{Z}_{2}$ reflection will only %%@
increase the amplitude in the SM throat by
a factor of 2 (i.e. constructive interference), and when it tunnels back through the barrier will contribute a %%@
vanishingly small amount to the I throat.  
Performing the calculation explicitly by including the reflected wave confirms this conclusion (see Appendix A), %%@
namely that the probability Eq.(\ref{eq:TunnProb})
is effectively unchanged.

The tunneling rate from throat I to throat SM is found by multiplying the tunneling probability Eq.(\ref{eq:TunnProb}) %%@
by the flux, the inverse
of the potential wall length $\frac{1}{|z_{i}|}$, so the tunneling rate from throat I to throat SM is:
\begin{equation}
\Gamma_{tunn} \sim (mR)^{4} \frac{1}{|z_{i}|}.
\label{eq:TunnRate}
\end{equation}
Note that when a KK mode tunnels from throat I to throat SM the mass $m$ of the mode stays (approximately) the same, %%@
and since the $AdS$ lengths are the same, the
tunneling probability back to the I throat is the same (making the $AdS$ lengths different will make the tunneling %%@
probability more asymetrical, suppressing the 
tunneling more in one direction than the other).  The tunneling rate, however, is inversely proportional to the %%@
(conformal) length of the throat, thus
{\it modes in a long throat take longer to tunnel out}.  Note that the depth of the two wells is approximately the %%@
same and does not cause much difference in the tunneling rates.  
This can be understood semi-classically as follows.  The average time required for a particle to 
escape a potential well is the average number of collisions required to escape multiplied by the time between %%@
collisions.  The average number of collisions
required is the inverse of the tunneling probability, $P^{-1}$.  The time required is twice the length of the well %%@
divided by the speed of the particle.  
In conformal coordinates, Eq.(\ref{eq:conformalMetric}), the KK modes behave like massless Klein-Gordon particles, and %%@
the length of the potential well is measured
in terms of the conformal length, thus the average time required for a KK mode to escape is $\tau_{tunn}\sim %%@
P^{-1}2|z_{i}|$, which gives the tunneling rate
Eq.(\ref{eq:TunnRate}) (up to a factor of two).

The tunneling rate can also be represented as the mixing between two weakly coupled potential wells, as in %%@
\cite{Kofman}:
\begin{equation}
\Gamma = |\bra{SM}e^{-iH\Delta t}\ket{I}|^{2}\cdot \mbox{Flux},
\label{eq:PropagatorGamma}
\end{equation}
where
\begin{equation}
H = \left(\begin{array}{cc} m & \epsilon \\ \epsilon & m' \end{array}\right)
\end{equation}
is the Hamiltonian for the coupled system, $\Delta m = |m-m'|$, and $\epsilon$ is the coupling of the systems.  Note %%@
that $\Delta m$ is the difference
in mass between states in the two throats, not the mass gap within the throats.  To leading order in $\epsilon$ one %%@
finds,
\begin{equation}
\Gamma \approx \frac{\epsilon^{2}}{\Delta m^{2}}\cdot \mbox{Flux}.
\label{eq:KofmanTunnRate}
\end{equation}
The coupling of the two systems $\epsilon$ should be proportional to some power of the tunneling probability; %%@
\cite{Kofman} suggests that
$\epsilon\sim m (mR_{-})^{2}$, which seems to imply $\Gamma \sim \frac{m^{2}}{\Delta m^{2}} (mR_{-})^{4}\cdot %%@
\mbox{Flux}$; in order to get Eq.(\ref{eq:TunnRate})
we must have $m\sim \Delta m$.  But this says that the difference in mass between the states in different wells is %%@
proportional to the mass level, which seems
unlikely in a general setup.
Instead, one can show that for a symmetric double well,
the expression for $\epsilon$ should be $\epsilon \sim \Delta m (mR_{-})^{2}$ (note
that the flux is still $z_{i}^{-1}$). This makes sense, since the coupling between two weakly coupled potential wells %%@
should be proportional to the splitting of
their adjacent masses.

A crucial aspect (as noted in \cite{Reheating}) of the viability of warped reheating is the ability to channel energy %%@
into Standard Model degrees of freedom and
not to lose energy to bulk gravitons, which can ruin BBN and other cosmological observations.  Naively one would %%@
expect the perturbative decay of KK modes to gravitons
to dominate over the non-perturbative tunneling effects\footnote{We thank D. Chung for emphasizing this point to us.}, 
thus reheating would fail to channel energy onto the Standard Model branes effectively.  However, several effects seem %%@
to render this intuition invalid.  First,
as noted above, the warping induces small couplings between the gravitons and KK modes.  Second, KK to graviton decay %%@
must occur via pair annihilation in order
to conserve extra dimensional quantum numbers, so the decay can only happen when a KK mode finds its partner with %%@
opposite quantum numbers.  Because of these
effects, it is possible that there exists a range of parameters for the model for with %%@
$\frac{\tau_{tunn}}{\tau_{g_{\mu\nu}}}\lesssim 1$.
Since thermalization drops the KK states down to the lowest mass levels, we will take $m\sim h_{i}/R_{-}$, 
which is the longest timescale for tunneling.  We find,
\begin{equation}
\frac{\tau_{tunnel}}{\tau_{g_{\mu\nu}}}\sim \frac{g_{s}^{1/4}}{\lambda^{11}}\left(\frac{M_{Pl}}{M_{i}}\right)^{2}.
\label{eq:ITunnDecay}
\end{equation}
We find that for,
\begin{equation}
\lambda\geq\left(\left(\frac{M_{Pl}}{M_{i}}\right)^{2}g_{s}^{1/4}\right)^{1/11},
\end{equation}
one can suppress the decay to gravitons.  For $M_{i}/M_{Pl}\sim 10^{-4}$ and $g_{s}\sim .1$ this gives a lower limit %%@
$\lambda\geq 5$, and keeps $R_{-}>\ell_{s}$,
which is what we wanted in order to trust the SUGRA approximation anyway.  However, this sets $h_{i}\sim 10^{-2}$, so %%@
the ``throat'' is not very strongly
warped, which also helps the modes tunnel through.

An important question is whether the tunneling takes place within the Hubble time after inflation, for otherwise the %%@
tunneling does not take place until the 
Hubble rate falls below the tunneling rate (which has implications for KK interactions in the SM throat).  The Hubble %%@
time after inflation is found from 
$H_{I}^{2} = \frac{1}{M_{Pl}^{2}}\epsilon_{0}$, and so
the ratio of the tunneling time to the Hubble time is,
\begin{equation}
\frac{\tau_{tunnel}}{H_{I}^{-1}} \sim \frac{g_{s}^{3/4}}{\lambda^{11}}\left(\frac{M_{Pl}}{M_{i}}\right)^{3}.
\end{equation}
Thus, for $\lambda \lesssim \left(\left(\frac{M_{Pl}}{M_{i}}\right)^{3}g_{s}^{3/4}\right)^{1/11}\sim 10$ 
we must consider Hubble
expansion for tunneling.  Note that while this gives us a small range of $\lambda$, this small range already %%@
implicitly existed since $h_{i}\propto \lambda^{3}$ and
we want $h_{i}<e^{-1}$ for a warped throat, thus too large of $\lambda$ ruins the warping of the throat.  One can also %%@
check that the KK modes thermalize 
within a Hubble time for $g_{s}\geq (M_{Pl}/M_{i})^{2}\sim 10^{-8}$ 
and so
for $\lambda$ within the values 
$\left(\left(\frac{M_{Pl}}{M_{i}}\right)^{2}g_{s}^{1/4}\right)^{1/11}\lesssim \lambda %%@
\lesssim\left(\left(\frac{M_{Pl}}{M_{i}}\right)^{3}g_{s}^{3/4}\right)^{1/11} $ 
($5\lesssim \lambda \lesssim 10$), corresponding to $1.9~\ell_{s} \lesssim R_{-} \lesssim 4.5~\ell_{s}$ (which are %%@
within our approximation $R_{-}>\ell_{s}$), we have the hierarchy of timescales: 
$\tau_{therm}\ll H_{I}^{-1}\lesssim \tau_{tunnel}\lesssim \tau_{g_{\mu\nu}}$.

\section{Standard Model Throat}
\setcounter{equation}{0}

After the KK modes from the I throat tunnel into KK modes in the SM throat these KK modes can do several things
in the SM throat, each with its own timescale: 1) decay to gravitons $\Delta t_{g_{\mu\nu}}$, 2) decay to Standard %%@
Model degrees of freedom $\Delta t_{SM}$, 3)
tunnel out of the SM throat $\Delta t_{tunnel}$ and 4) thermalize and drop down to relativistic degrees of freedom at %%@
the bottom of the throat $\Delta t_{therm}$, all
of which can happen relative to 5) the Hubble timescale $H_{tunn}^{-1}$.

Before we determine any of these timescales, we must determine the approximate number density of the KK modes after %%@
tunneling, which will be important in our
calculation of the thermalization and graviton decay timescales.  Since the number density falls off as $a(t)^{-3}$, %%@
where $a(t)$ is the scale factor, the 
number density after tunneling is,
\begin{equation}
n_{tunn}\sim n_{I}\left(\frac{a_{I}}{a_{tunn}}\right)^{3},
\label{eq:NdensityEvolve}
\end{equation}
We will estimate the energy density for the modes to be non-relativistic, which falls off as $a^{-3}$, and since the %%@
tunneling will happen after the Hubble rate drops below
the tunneling rate, i.e. $H^{2}_{tunn}\sim \frac{1}{\tau_{tunn}^{2}}\sim \frac{1}{M_{Pl}^{2}}\epsilon_{tunn}$ we have:
\begin{equation}
\epsilon_{tunn} \sim \frac{M_{Pl}^{2}}{\tau_{tunn}^{2}} \sim \epsilon_{I}\left(\frac{a_{I}}{a_{tunn}}\right)^{3}.
\label{eq:SecondRatio}
\end{equation}
Putting Eqs.(\ref{eq:NdensityEvolve})(\ref{eq:SecondRatio}) together we find
\begin{equation}
n_{tunn}\sim n_{I}\left(\frac{H_{I}^{-1}}{\tau_{tunn}}\right)^{2}\sim %%@
\frac{M_{i}^{3}}{g_{s}}\left(\frac{H_{I}^{-1}}{\tau_{tunn}}\right)^{2}.
\label{eq:NumbDensTunn}
\end{equation}
We can now use this to estimate the thermalization and graviton decay time scales 
for the KK modes in the SM throat as in Section 3.

\subsection{KK graviton decay}

We again estimate the decay of KK modes to gravitons as $\Delta t_{g_{\mu\nu}}\sim (n\sigma v)^{-1}$, where we will %%@
take $v\sim 1$ and $\sigma\sim M_{Pl}^{-2}$
as before.  For the number density we will use the estimate Eq.(\ref{eq:NumbDensTunn}) after tunneling, and our %%@
estimate for the graviton decay timescale is then
\begin{equation}
\Delta t_{g_{\mu\nu}} \sim %%@
\left(\frac{M_{Pl}}{M_{i}}\right)^{2}\left(\frac{\tau_{tunn}}{H_{I}^{-1}}\right)^{2}\frac{g_{s}}{M_{i}}.
\label{eq:SMGravDecay}
\end{equation}

Similarly, we estimate the timescale for thermalization $\Delta t_{therm}$ of the KK modes where we note that now the %%@
cross-section is suppressed only by the
local string scale in the SM throat, $\sigma \sim g_{s}^{2} (M_{s}h_{sm})^{-2}=g_{s}^{2}M_{sm}^{-2}$, so that the %%@
thermalization time scale is again much faster than the decay
to gravitons, and is given by:
\begin{equation}
\Delta t_{therm} \sim %%@
\left(\frac{M_{sm}}{M_{i}}\right)^{2}\left(\frac{\tau_{tunn}}{H_{I}^{-1}}\right)^{2}\frac{1}{g_{s} M_{i}}.
\label{eq:SMTherm}
\end{equation}

\subsection{Decay to Standard Model degrees of freedom}

The decay rate of KK particles to Standard model degrees of freedom on a brane located at $y=y_{0}$ in the throat can %%@
be estimated by considering the interaction
to be of the form \cite{SMCoupling1, SMCoupling2, SMCoupling3} (where we are considering only the $AdS_{5}$ part of %%@
the throat here):
\begin{equation}
{\cal L}_{int} \sim -\frac{1}{M_{Pl,5}^{3/2}}T^{\mu\nu}(x) h_{\mu\nu}(x,y_{0}),
\label{eq:KKSML}
\end{equation}
where $M_{Pl,5}$ is the 5-D Planck mass, related to the 4-D Planck mass by $M_{Pl}^{2} = M_{Pl,5}^{3}R %%@
(1-h_{sm})\approx M_{Pl,5}^{3}R$, 
and $T^{\mu\nu}(x)$ is the energy-momentum
tensor of the Standard Model fields.  The metric flucuation $h_{\mu\nu}(x,y)$ can be expanded as:
\begin{equation}
h_{\mu\nu}(x,y) = \sum_{n=0}^{\infty} h_{\mu\nu}^{(n)}(x)\frac{\psi^{(n)}(y)}{\sqrt{R}}.
\label{eq:KKexpand}
\end{equation}
The 5-D part of the wavefunction $\psi^{(n)}(y)$ is given by the general formula Eq.(\ref{eq:generalsoln1}) and is %%@
normalized as
$\int dy e^{y/R} \psi^{(m)}(y)\psi^{(n)}(y) = \delta_{nm}$.  Using the normalization, the expansion of the KK modes %%@
Eq.(\ref{eq:KKexpand}), and the definition
of the 5-D Planck mass we have then,
\begin{equation}
{\cal L}_{int} = -\frac{1}{M_{Pl}}T^{\mu\nu}(x)h^{(0)}_{\mu\nu}(x) - %%@
\frac{1}{h_{sm}M_{Pl}}T^{\mu\nu}(x)\sum_{n=1}^{\infty} h^{(n)}_{\mu\nu}(x),
\label{eq:KKSMSuppres}
\end{equation}
where $h_{sm}$ is set by the local string scale $M_{sm}$ at the tip of the Standard Model throat.

The decay rate to Standard Model fields is then approximately \cite{SMCoupling2}:
\begin{equation}
\Gamma_{SM} \sim m_{KK}^{3} \frac{1}{M_{sm}^{2}}.
\label{eq:KKSMRate}
\end{equation}
A more precise determination of the decay rate, including loop corrections, can be found in \cite{SMCoupling1, %%@
SMCoupling2, SMCoupling3, KKBrane}, where we have
ignored the extra contributions which depend on the ratio of the mass of the Standard Model particles and the mass of %%@
the KK modes, $m_{sm}/m_{KK}\ll 1$.  
Note that this estimate is similar to that of \cite{Kofman}, where they consider decay of KK modes into the scalar %%@
transverse excitations of the brane 
\cite{KKBrane} which then quickly decay into Standard Model degrees of freedom:
\begin{equation}
\Gamma_{KK}\sim  m_{KK}^{3}\frac{\sqrt{1-\frac{4 \mu^{2}}{m_{KK}^{2}}}}{M_{sm}^{2}}.
\label{eq:KKSMBrane}
\end{equation}
Here $\mu$ is the effective mass of the $\overline{D3}$ ($D3$) induced by the background fluxes (SUSY breaking).  We %%@
see that for 
$\mu\ll m_{KK}$ we can neglect the square root in Eq.(\ref{eq:KKSMBrane}) and this decay rate is similar to the decay %%@
to Standard Model particles.  
However, since $m_{KK}\sim h_{sm}R_{-}^{-1}$ 
and a $\overline{D3}$ will have $\mu\sim h_{sm}/R_{-}$ from the fluxes, 
the decay rate to the transverse fluctuations becomes effectively zero.  Decays directly to Standard Model degrees of %%@
freedom thus have more phase space 
and we will use Eq.(\ref{eq:KKSMRate}) to determine the decay time to Standard Model 
degrees of freedom.  Using $m\sim h_{sm}R_{-}^{-1}$ we find,
\begin{equation}
\Delta t_{SM}\sim \frac{\lambda^{3}g_{s}^{3/4}}{M_{sm}(\ln h_{sm}^{-1})^{3/4}}
\label{eq:SMDecayTime}
\end{equation}

\subsection{Reheating}

We would like the decay of KK modes into Standard Model particles Eq.(\ref{eq:SMDecayTime}) to be quicker than the %%@
decay into gravitons 
Eq.(\ref{eq:SMGravDecay}),
\begin{equation}
\frac{\Delta t_{sm}}{\Delta t_{g_{\mu\nu}}}\sim \lambda^{25}\left(\frac{M_{i}}{M_{Pl}}\right)^{8}\frac{M_{i}}{M_{sm}} 
\frac{(\ln h_{i}^{-1})^{1/2}}{g_{s}^{7/4}(\ln h_{sm}^{-1})^{3/4}}.
\end{equation}
We find that for $M_{sm}\sim M_{i}$, this ratio is less than one for the range of $\lambda$ quoted at the end of %%@
Section 3, thus the decay to Standard Model
degrees of freedom is favored.  However, for $M_{sm}/M_{i}\sim 10^{-11}$ (i.e. $M_{sm}\sim$ Tev if one wants to %%@
generate the weak scale hierarchy during inflation)
we must bring $\lambda$ down to such small values in order to have this process be favored that it does not appear we %%@
can simultaneously ensure that the KK modes
in the inflationary throat tunnel and the KK modes in the SM throat reheat the Standard Model degrees of freedom.  %%@
This can be seen by noting that less massive
modes couple weaker to Standard Model degrees of freedom, Eq.(\ref{eq:KKSMRate}).  A deeper throat means that the %%@
lightest KK modes in the throat are less
massive, thus couple weaker to the Standard Model.  Meanwhile $\Delta t_{g_{\mu\nu}}$ is insensitive to the details of %%@
the throat construction, so eventually 
KK to graviton decay is favored.

The ratio of the time to decay to Standard Model particles and the Hubble time is,
\begin{equation}
\frac{\Delta t_{SM}}{H_{tunn}^{-1}} \frac{M_{i}}{M_{sm}}\left(\frac{M_{i}}{M_{Pl}}\right)^{4}\frac{1}{(\ln %%@
h_{sm}^{-1})^{1/4}}
\frac{\lambda^{14}}{g_{s}^{1/2}}
\end{equation}
This ratio is less than one for $M_{i}\sim M_{sm}$ as we took above and for all $\lambda$ in the range quoted above at %%@
the end of Section 3.  Thus, we expect
that the reheating of the Standard Model from KK modes decaying in the Standard Model throat will take place %%@
immediately after those modes tunnel.  Also
note that $\Delta t_{sm}< H_{I}^{-1}\approx \tau_{tunn}\leq \Delta t_{tunn}$, with the last inequality saturated for %%@
$M_{i}\sim M_{sm}$, so the modes will
not tunnel back out before decaying to the Standard Model.  One can also check that the ratio of the thermal time to %%@
the Hubble scale after tunneling is,
\begin{equation}
\frac{\Delta t_{therm}}{H_{tunn}^{-1}} \approx %%@
\left(\frac{M_{sm}}{M_{i}}\right)^{2}\left(\frac{M_{Pl}}{M_{i}}\right)^{2}\frac{1}{\lambda^{7} g_{s}^{3/4}},
\end{equation}
which is greater than one for $M_{sm}\sim M_{i}$ and the appropriate ranges of $\lambda$.  Thus, the KK modes are no %%@
longer thermal once they tunnel into the 
Standard Model throat.

Because the decay of the KK modes in the Standard Model throat is very quick, reheating happens when the Hubble rate %%@
drops below the tunneling rate, 
Eq.(\ref{eq:TunnRate}); when this happens the Standard Model brane is reheated and the energy density is
\begin{equation}
\epsilon_{RH} \sim g_{*}T_{RH}^{4} \sim M_{Pl}^{2}H_{tunn}^{2} \sim \frac{M_{Pl}^{2}}{\tau_{tunn}^{2}}.
\end{equation}
Here $g_{*}\sim 100$ is the number of Standard Model degrees of freedom, and solving for the reheating temperature we %%@
obtain:
\begin{equation}
T_{RH}\sim %%@
\frac{M_{i}}{g_{*}^{1/4}}\left(\frac{M_{i}}{M_{Pl}}\right)^{3/2}\left[\frac{\lambda^{11}}{g_{s}^{5/4}}\right]^{1/2}.
\label{eq:ReHeat}
\end{equation}

For $M_{i}\sim M_{sm}$ and the range of values of $\lambda$ for which the reheating process seems to efficiently %%@
channel energy onto the Standard Model brane, i.e. for the tunneling
out of the I throat to be quicker than decay into gravitons and for the decay to Standard Model particles to be faster %%@
than the decay to gravitons in the SM throat
we have the range \footnote{This range of reheating temperatures depends on our choice of the local string scale at %%@
the inflationary throat. One could obtain lower reheating temperature if the local string scale is not set to 
generate the precise level of density perturbations as we do here.} 
of reheating temperatures $10^{13} GeV\leq T_{RH}\leq 10^{14} GeV$.  Notice that this range of values is smaller than %%@
$M_{sm}\sim 10^{15} GeV$, which
implies that the KK modes {\it do not} excite stringy degrees of freedom, and we can trust this field-theory %%@
description of reheating.  
Note also that larger values
of $\lambda$ could push the reheat temperature above $M_{i}$, but then we would have little to no warping for the %%@
inflationary throat (see Eq.(\ref{eq:WarpLambda})).  

\section{Adding More Throats}
\setcounter{equation}{0}

In a scenario with multiple (more than two) throats there are additional concerns that must be addressed: How are the %%@
tunneling KK modes from the I throat partitioned
between the throats?  Will KK modes from other throats decay to gravitons or decay to Standard Model particles at late %%@
times, ruining BBN?

To address the first question, notice that Eq.(\ref{eq:TunnRate}) for the tunneling rate of KK modes from the I throat %%@
does not depend at all on the details of the
SM throat.  One can make this more explicit by writing the mass as $m_{n}\sim \frac{n h_{i}}{R_{-}}$ and $z_{i}\sim %%@
R_{+}h_{i}^{-1}$: $\Gamma_{tunn}\sim n^{4}h_{i}^{5}/R_{+}$.  
Thus we expect that each throat will receive equal amounts of KK modes from the tunneling process.  A similar %%@
conclusion was found in \cite{Kofman} by 
modifying the earlier discussion of the tunneling rate in Section 3, using the Hamiltonian.  
However, the 4-D mass of the KK
modes is conserved (actually, it may be broken by the gluing of the throat to the CY 3-fold bulk
or the presence of D-branes, but this breaking should be further suppressed) so a KK mode
of mass $m_{n}\sim \frac{n h_{i}}{R_{-}}$ in the I throat can only decay into a mode with mass $m_{n}\approx %%@
m_{n'}\sim \frac{n' h_{X}}{R_{-}}$ in another throat with 
warp factor $h_{X}$.  For a long throat like the SM throat the mass spacing $\Delta m\sim \frac{h_{sm}}{R_{-}}$ is %%@
small so it is more likely for a tunneling mode
to find a (nearly) equal mass in this throat than in less-warped throats (this is also true for modes tunneling from a %%@
throat other than I).  However,
it has been suggested that stringy corrections will 
shorten a
long throat to be approximately the same length as the inflationary throat (more precisely, the
length is set by the Hubble scale during inflation) \cite{Frey:2005jk};
in this case the lengths of all throats will be approximately the same, 
and the KK modes will again be able to find equal mass states to tunnel into.  

If there are other throats with the same length as the I throat (modified in the same way due to string corrections as %%@
the SM throat) then one would expect these
throats to also receive a significant amount of energy from the $D\overline{D}$ decay.  However, since after %%@
tunneling, $H < \Gamma_{tunn}$, the KK modes in these other throats
will quickly decay out of these throats, finding their way into the SM throat, as long as there are no branes at the %%@
tips of the other throats for the KK modes to decay
onto.  If the length of these throats is changing, however (as may happen in the relaxation of the deformation modulus %%@
after inflation), they could trap the KK modes
as the potential well deepens.  However, in this case one should also be concerned with the particle production from %%@
the relaxation of the deformation modulus to its 
ground state \cite{Frey:2005jk}, which deserves further study.

If we can somehow sidestep the issues raised in \cite{Frey:2005jk} and 
build multi-throat inflationary models with widely different warping in a stable and consistent way,
then we should be concerned about the tunneling of KK modes from these other ``X'' throats into the SM throat, leading %%@
to late-time reheating.  We should require that any such
reheating must occur (roughly) above $\sim 10\ MeV$, where it can conflict with BBN.  Again, the reheating from the %%@
other throats will happen when the tunneling rate drops below 
the Hubble rate, which will lead to a reheating temperature (assuming a significant fraction of energy was transfered %%@
to the ``X'' throat from the tunneling from the 
inflationary throat):
\begin{equation}
T_{RH} = %%@
\frac{M_{X}}{g_{*}^{1/4}}\left(\frac{M_{X}}{M_{Pl}}\right)^{3/2}\left[\frac{\lambda^{11}}{g_{s}^{5/4}}\right]^{1/2}.
\end{equation}
Keeping this reheating temperature above $10\ MeV$ requires for $\lambda\sim 5$ ($\lambda\sim 10$) that the local %%@
string scale in the other throat is 
$M_{X}\geq 10^{9}\ GeV$ ($M_{X} \geq 10^{8}\ GeV$), which corresponds to a warp factor of $h_{X}\geq10^{-8}$ %%@
($h_{X}\geq 10^{-7}$).

\section{Conclusion}

In this paper, we 
discuss the issue of warped reheating in  multi-throat brane inflation.
There are several added appealing features in
brane inflationary models with more than one throat, the most exciting of which is perhaps the prospect of producing %%@
long-lived cosmic superstrings
as a window to see stringy physics \cite{CosStrings,Cosmic,FandD,CosmicStringReviews}.
For this scenario to hold up, however, it is essential to establish that 
the Standard Model can be successfully reheated.
In particular, irrespective of how energy is ultimately deposited to the Standard Model, whether via a conventional
approach \cite{Reheating,Kofman} or more stringy effects such as open string decay \cite{Frey:2005jk}, one needs to %%@
first demonstrate that the 
energy in the inflationary throat can be effectively channeled to the throat
where the Standard Model lives. At first sight, this seems to pose a challenge since
 the KK modes will naively decay predominantly into gravitons.
However, we found that over a modest range of parameters of 
the background geometry, the KK modes will preferably tunnel out of the inflationary throat
instead. 
This can be understood as the result of several combined effects:
(i) First of all, the coupling of the KK modes to gravitons are suppressed by warping
since the graviton wavefunction is exponentially suppressed at the tip of the throat. 
(ii) The number density of gravitons is also diluted somewhat by warp factor of
the inflationary throat.
(iii) The compactification scale can be slightly larger than the string scale (which we assume 
anyway in order to trust a SUGRA description of flux compactification). Therefore, 
only a mild warp factor (say $h \sim 10^{-2}$) is needed for the inflationary throat
and the potential well from which the KK modes have to tunnel out is not very steep after all.  We also investigate %%@
the decay of closed strings 
and find that the decay into KK modes is preferred in a warped throat due to the warp factor enhancement of the %%@
wavefunction.

Note that for simplicity, we consider throats with the same AdS curvature
scale in our analysis. This is not a requirement in the underlying
flux compactification of string theory, and  
we expect that the constraints presented here can be more easily satisfied by allowing
the different throats to have different AdS curvature.  One could also consider relaxing
the constraints on the local string scale at the inflationary throat if the smallness of the observed
density perturbation comes from other sources, such as features of the inflaton potential and/or
other mechanisms of generating fluctuations.  
We leave this and other straightforward extensions for future work.

Clearly, this and the earlier works \cite{Reheating,Kofman,Frey:2005jk} are
some initial forays into the problem of reheating in brane inflation.
More work is needed to substantiate this picture, 
including a better understanding of the decay of long open strings in \cite{Frey:2005jk}
and the issue of long-lived angular KK modes \cite{Kofman}.
The latter for instance impose non-trivial constraints on the isometries of the throats.
Turning the problems around, such considerations may provide,
in addition to \cite{Brandenberger,Shiu,Danielsson,Kempf,Kaloper,Burgess,BEFT,Porrati,CH,Mottola}, yet other powerful
ways of probing string scale physics.

\section{Acknowledgements}
We would like to thank Cliff Burgess, Xin-gang Chen, Daniel Chung, Hooman Davoudiasl, Min-xin Huang,
Lev Kofman, Peter Langfelder, Rob Myers, and Piljin Yi for discussions.  
GS and BU would also like to thank the Perimeter Institute for Theoretical Physics for hospitality while part of this %%@
work was done.  The work of 
GS and BU was supported in part by NSF CAREER Award No. PHY-0348093, DOE grant DE-FG-02-95ER40896, a Research %%@
Innovation Award and a Cottrell 
Scholar Award from Research Corporation.  DC acknowledges partial
support by the EC-TRN network MRTN-CT-2004-005104 and also by the
Italian MIUR program ``Teoria dei Campi, Superstringhe e Gravita'' and
by INFN sezione M12.

\appendix

\section{Tunneling with reflection}
\setcounter{equation}{0}

Consider a background with two Randall-Sundrum (RS) throats glued together at the Planck brane, where the two throats %%@
have different warp factors, with metric,
\begin{equation}
ds^{2}=\frac{R^{2}}{(|z|+R)^{2}}(\eta_{\mu\nu}dx^{\mu}dx^{\nu}+dz^{2}),
\end{equation}
where $R$ is the AdS length, and $-l_{1}\leq z \leq l_{2}$.  The utility of expressing the metric in these coordinates %%@
is that the equation of motion for the KK modes of the graviton,
\begin{equation}
h_{\mu\nu}(x,z)=\sqrt{\frac{R}{|z|+R}}e^{ipx}\psi_{\mu\nu}(z),
\label{eq:KKmode2}
\end{equation}
becomes a simple Schr\"odinger equation for $\psi$ (suppressing the indicies):
\begin{equation}
\partial_{z}^{2} \psi (z)+(m^{2}-\frac{15}{4(|z|+R)^{2}})\psi(z) = 0,
\label{eq:eqofmotion2}
\end{equation}
where the potential felt by the KK mode is proportional to the square of the warp factor, see Figure %%@
\ref{fig:2ThroatsPotential}; 
here $p^{2}=m^{2}$ is the mass of the KK mode.

The general solution to this equation of motion can be expressed in terms of Bessel functions,
\begin{equation}
\psi(z)=\sqrt{m(|z|+R)}[A J_{2}(m(|z|+R))+B Y_{2}(m(|z|+R))].
\label{eq:generalsoln2}
\end{equation}

We would like to consider tunneling of an incoming state from the left hand throat into a state on the right hand %%@
throat as in \cite{Small}, however, one might be concerned that \cite{Small} have neglected the termination of the %%@
right hand throat  by not including a ``reflected" term in the SM throat.  Let us consider this case more carefully by %%@
considering wavefunctions,
\begin{eqnarray}
\psi_{1}(z) &=& \sqrt{m(|z|+R)}A [H_{2}^{(1)}(m(|z|+R))+B H_{2}^{(2)}(m(|z|+R))] \nonumber \\
\psi_{1}(z) &=& \sqrt{m(|z|+R)}C [H_{2}^{(1)}(m(|z|+R))+D H_{2}^{(2)}(m(|z|+R))].
\end{eqnarray}
The matching and jump conditions at the Planck brane are
\begin{eqnarray}
\psi_{1}(0) &=& \psi_{2}(0)~, \nonumber \\
\partial_{z}\psi_{2}|_{z=0}-\partial_{z}\psi_{1}|_{z=0} &=& -\frac{3}{R}\psi_{1}(0),
\label{eq:BndConditions}
\end{eqnarray}
which give (the suppressed argument is $mR$):
\begin{eqnarray}
A(H_{2}^{(1)}+B H_{2}^{(2)}) &=& C(H_{2}^{(2)}+DH_{2}^{(2)})~, \nonumber \\
A(H_{1}^{(1)}+B H_{1}^{(2)}) &=& -C(H_{1}^{(2)}+DH_{1}^{(2)}).
\label{eq:MatchJump}
\end{eqnarray}
The boundary conditions at $z=-l_{1}$ and $z=l_{2}$ are
\begin{eqnarray}
\partial_{z}\psi_{1}|_{z=-l_{1}} = \frac{3}{2(l_{1}+R)} \psi_{1}(-l_{1})~,\nonumber \\
\partial_{z}\psi_{2}|_{z=l_{2}} = -\frac{3}{2(l_{2}+R)} \psi_{2}(l_{2}).
\label{eq:IRBndCond}
\end{eqnarray}
Notice that these boundary conditions translate into,
\begin{eqnarray}
\chi_{i}(0)&=&\chi_{sm}(0) \nonumber \\
%\frac{d}{dz}[(|z|+L)^{2} {\cal Z}_{i}(z)]|_{z=0} &=& -\frac{d}{dz}[(|z|+L)^{2} {\cal Z}_{sm}(z)]|_{z=0} %\nonumber \\
%\rightarrow 
\frac{d}{dy}\chi_{i}(y)|_{y=0} &=& -\frac{d}{dy}\chi_{sm}(y)|_{y=0} \nonumber \\
\frac{d}{dy}\chi_{i}(y)|_{y=-y_{i}} &=& 0 \nonumber \\
\frac{d}{dy}\chi_{sm} (y)|_{y=y_{sm}} &=& 0
\end{eqnarray}
where 
%${\cal Z}$ represents a linear combination of Bessel functions, 
$|z|+R=R e^{k|y|}$, 
${\cal Z}$ represents a linear combination of Bessel functions, 
and $\chi(y)=e^{2k|y|}{\cal Z}$ 
is the wavefunction for the
KK modes in Eq.(\ref{eq:BesselSoln}).

We recognize $C$ as the amplitude of the outgoing wave in the SM throat, while $AB$ is the incoming wave in I throat, %%@
$A$ is the reflected wave in the I throat, and $CD$ is the reflected wave in the SM throat.  The SM reflected wave %%@
looks like an incoming source for the scattering problem, thus the probability of tunneling is given by,
\begin{equation}
P=\left|\frac{C}{AB+CD}\right|^{2} = \left|\frac{1}{\frac{AB}{C}+D}\right|^{2},
\label{eq:TunnelProb}
\end{equation}
and the probability of reflection is,
\begin{equation}
R=\left|\frac{A}{AB+CD}\right|^{2} = \left|\frac{1}{B+\frac{C}{A}D}\right|^{2}.
\end{equation}

The I throat boundary condition Eq.(\ref{eq:IRBndCond}) again sets the mass gap for the problem, $mR\sim n a_{i}$, so %%@
for low modes $mR\ll 1$ (with $m l_{2}\gg 1$ i.e. $1\ll n\ll h_{i}^{-1}$) we have,
\begin{eqnarray}
P &\approx & %%@
\frac{16}{\pi^{2}(mR)^{2}}\left|\frac{1}{\frac{-16}{\pi^{2}(mR)^{3}}+\frac{-16}{\pi^{2}(mR)^{3}}}\right|^{2}=\frac{\pi%%@
%%@
^{2}}{64}(mR)^{4} \nonumber \\
R &\approx & 1
\label{eq:BoundProb}
\end{eqnarray}

Notice that Eqs.(\ref{eq:TunnProb})(\ref{eq:BoundProb}) give nearly identical tunneling and reflection probabilities.  %%@
We can understand this by noticing that adding a SM boundary condition reflects the transmitted wave back (giving a %%@
slightly larger ``incoming" flux from the right), but when this reflection tunnels through the barrier it is %%@
suppressed again by $\sim (mR)^{4}$, thus its effect on the original incoming wave in the I throat is negligible, so %%@
the scattering problem is not changed much.

\end{document}